\documentclass[a4paper,twocolumn,showpacs]{revtex4}%
\usepackage{mathrsfs}
\usepackage{amsfonts}
\usepackage{amsmath}
\usepackage{amssymb}
\usepackage{graphicx}
\usepackage{float}%
\providecommand{\U}[1]{\protect\rule{.1in}{.1in}}
\begin{document}
\title{Artificial Gauge Field and Quantum Spin Hall States in a Conventional
Two-dimensional Electron Gas}
\author{Likun Shi$^{1}$, Wenkai Lou$^{1}$, F. Cheng$^{1}$, Y. L. Zou$^{1}$, Wen
Yang$^{2}$ and Kai Chang$^{1}$}
\affiliation{$^{1}$SKLSM, Institute of Semiconductors, Chinese Academy of Sciences, P.O.
Box 912, Beijing 100083, China}
\affiliation{$^{2}$Beijing Computational Science Research Center, Beijing 100094, China}

\begin{abstract}
Based on the Born-Oppemheimer approximation, we divide total electron
Hamiltonian in a spin-orbit coupled system into slow orbital motion and fast
interband transition process. We find that the fast motion induces a gauge
field on slow orbital motion, perpendicular to electron momentum, inducing a
topological phase. From this general designing principle, we present a theory
for generating artificial gauge field and topological phase in a conventional
two-dimensional electron gas embedded in parabolically graded GaAs/In$_{x}%
$Ga$_{1-x}$As/GaAs quantum wells with antidot lattices. By tuning the etching
depth and period of antidot lattices, the band folding caused by superimposed
potential leads to formation of minibands and band inversions between the
neighboring subbands. The intersubband spin-orbit interaction opens
considerably large nontrivial minigaps and leads to many pairs of helical edge
states in these gaps.

\end{abstract}

\maketitle

\section{INTRODUCTION}

Exploring of various topological quantum states is always one of the central
issue of condensed matter physics \cite{TQS1,TQS2,TQS3}. Topological
insulators (TIs) \cite{RMP1}, a new class of solids, posses unique properties
such as robust gapless helical edge or surface states and exotic topological
excitations
\cite{RMP1,RMP2,Kane,BHZ,Konig1,Konig2,Fu,Fu2,Hsieh,ZXShen,Hasan,Lin,Franz,KYang,Chadov,Heusler,Chang1,Chang2,Chang3,GaAs,Oxide,Tinfilm,
Franz2,Katsnelson}. The helical edge states of two-dimensional (2D) TIs are
protected strictly against elastic backscattering from nonmagnetic impurities.
This feature leads to dissipationless conducting channels and therefore is
promising for possible applications in spintronics, quantum information,
thermoelectric transport and on-chip interconnection in integrated circuit.
These novel applications require large nontrivial gaps, which suppress the
coupling between the edge and bulk states, leading to dissipationless edge
transport. For this purpose, there is an ongoing search for feasible
realizations of various narrow gap materials containing heavy elements, e.g.,
CdTe/HgTe/CdTe quantum wells (QWs) \cite{BHZ,Konig1,Konig2}, and Tin film
\cite{Tinfilm}. However, fabrication of high-quality samples of these proposed
structures still remains a challenging task, requiring precise control for
material growth.

In this work, we demonstrate that conventional semiconductor GaAs/In$_{x}%
$Ga$_{1-x}$As/GaAs two-dimensional electron gas (2DEG) with antidot lattices
can be driven into the TI phase. The 2DEGs provide a promising playground for
realizing TI states with quite large nontrivial gap ($\sim20$ meV) operating
at liquid nitrigen temperature regime, instead of searching new materials
containing heavy atoms. We first present a general analysis for generating an
artificial gauge field in a semiconductor 2DEG, then we demonstrate band
inversion between neighboring subbands utilizing antidot lattices created by
well-developed semiconductor etching technique, and generate the TI phase with
many pairs of helical edge states. This suggests a completely new method to
generate topological phase in conventional semiconductor 2DEGs without strong
spin-orbit interaction (SOI), at liquid nitrigen temperature regime.

\section{GAUGE FIELD FROM BORN-OPPENHEIMER APPROXIMATION}

First we discuss the emergence of an artificial gauge field in a system of
electrons in a 2D system described by a low-energy single-particle Hamiltonian
$H={\epsilon}(\mathbf{k}){\sigma}_{0}\otimes\tau_{0}+\sum_{i,j=1}^{3}%
d_{ij}(\mathbf{k}){\sigma}_{i}\otimes\tau_{j}$, where ${\sigma}_{i}$ and
$\tau_{i}$ $(i=1,2,3)$ are Pauli matrices describing the electron spin and the
conduction ($\tau_{3}=+1$) and valence ($\tau_{3}=-1$) bands, respectively,
and $\sigma_{0}=\tau_{0}=I_{2\times2}$ are identity matrices. Taking
$d_{31}=Ak_{x}$, $d_{02}=Ak_{y}$, $d_{03}=M-Bk^{2}$ and other $d_{ij}=0$, we
obtain the Bernevig-Hughes-Zhang (BHZ) Hamiltonian for 2D TIs {\cite{BHZ}}.
Neglecting the band index $\tau$, and taking $d_{1}=-\alpha k_{y}-\beta k_{x}%
$, $d_{2}=\alpha k_{x}+\beta k_{y}$, $d_{3}=0$, we get the Hamiltonian for a
2DEG with Rashba and Dresselhaus SOIs, where $\alpha$ and $\beta$ are the
strengths of Rashba and Dresselhaus SOIs, respectively. Next, we divide the
total Hamiltonian into the intra-band, slow part $H_{\mathrm{orb}}={\epsilon
}(\mathbf{k}){\sigma}_{0}\otimes\tau_{0}$ and the inter-band, fast part
$H_{\mathrm{IB}}=\sum_{i,j=1}^{3}d_{ij}(\mathbf{k}){\sigma}_{i}\otimes\tau
_{j}$, which usually arises from the SOIs in real materials. The eigenstate of
the total Hamiltonian can be decomposed into the fast and slow components:
$|\Psi(\mathbf{k})\rangle=%
%TCIMACRO{\tsum \nolimits_{n}}%
%BeginExpansion
{\textstyle\sum\nolimits_{n}}
%EndExpansion
\phi_{n}(\mathbf{k})|\chi_{n}(\mathbf{k})\rangle$, where $\{|\chi
_{n}(\mathbf{k})\rangle\}$ are eigenstates of the fast part $H_{\mathrm{IB}}$
and $\{\phi_{n}(\mathbf{k})\}$ describe the slow part. The fast spin dynamics
compared with the slow orbital motion allow us to make the Born-Oppenhenmer
approximation, i.e., neglecting the coupling between different $|\chi
_{n}(\mathbf{k})\rangle$, and derive an effective Hamiltonian governing the
slow orbital motion $\phi_{n}(\mathbf{k})$ (see Appendix A):%
\begin{align}
H_{n}(\mathbf{k})  &  =\left\langle \chi_{n}(\mathbf{k})\left\vert
H_{\mathrm{orb}}(\mathbf{k},\mathbf{\hat{r}})+H_{\mathrm{IB}}(\mathbf{k}%
,\mathbf{\hat{\sigma}})\right\vert \chi_{n}(\mathbf{k})\right\rangle
\nonumber\\
&  =H_{\mathrm{orb}}(\mathbf{k},\mathbf{\hat{r}}-\mathbf{A}_{n})+\epsilon
_{n}(\mathbf{k}),
\end{align}
where $\epsilon_{n}(\mathbf{k})=$ $\left\langle \chi_{n}(\mathbf{k})\left\vert
H_{\mathrm{IB}}\right\vert \chi_{n}(\mathbf{k})\right\rangle $ acts as an
effective potential that seperates different bands $\{|\chi_{n}(\mathbf{k}%
)\rangle\}$ and $\mathbf{A}_{n}=-i\left\langle \chi_{n}(\mathbf{k})\left\vert
\partial_{\mathbf{k}}\right\vert \chi_{n}(\mathbf{k})\right\rangle $ is a
gauge potential in the momentum space of the slow orbital motion, due to the
interband coupling to the fast spin dynamics {\cite{Wilczek,CPSun}}. For the
BHZ Hamiltonian, the gauge potential $\mathbf{A}_{n}$ leads to an effective
Lorentz force $F_{xy,n}(\mathbf{k})$ in the momentum space perpendicular to
the electric field $\mathbf{E}$:%
\begin{align}
F_{xy,n}(\mathbf{k})  &  \equiv i[x^{\prime},y^{\prime}]=i[i\partial_{k_{x}%
}-A_{x,n},i\partial_{k_{y}}-A_{y,n}]\nonumber\\
&  =s_{n}t_{n}\frac{A^{2}(M+Bk^{2})}{2[A^{2}k^{2}+(M-Bk^{2})^{2}]^{3/2}},
\end{align}
where $s_{n}=\pm1$ denotes spin up or down state while $t_{n}=\pm1$ denotes
the conduction or valence band, respectively ($n=1,2,3,4$). The Chern number
$C_{n}=\left(  1/2\pi\right)  \int_{\mathrm{BZ}}\mathrm{d}^{2}kF_{xy,n}%
=s_{n}t_{n}\left[  1+\mathrm{sign}\left(  M/B\right)  \right]  /2$ is obtained
by integrating the field strength $F_{xy,n}$ in the Brillouin zone. The sign
change in $M$ would induce a change of the Chern number by 1, which
corresponds to the topological phase transition {\cite{Moore,Galitski}}.

For a 2DEG with Rashba and Dresselhaus SOIs, we find $F_{xy}(\mathbf{k})=0$,
which means that the Chern number vanishes in 2DEG with SOIs. Comparing the
Hamiltonian of 2D TIs to that of 2DEGs with SOIs, one can see clearly that
2D\ TIs posses an additional degree of freedom:\ the band index $\tau$. In
order to generate the gauge field and realize TI phases in a 2DEG, one needs
to create minibands and band inversion in 2DEGs.

\section{EFFECTIVE HAMILTONIAN AND QUANTUM SPIN HALL STATES}

Based on the above designing principle, we will create topological phase in
conventional semiconductor 2DEG. This is the first demonstration of the
formation of a TI phase in the \textit{s-like} band systems, i.e., a 2DEG with
nanostructured antidot lattice shown schematically in Fig. 1(a).
Nanostructured antidot lattices, consisting of periodically arranged holes
that are etched in a 2DEG, form a strongly repulsive egg-carton-like periodic
potential in a 2DEG
\cite{Antidot,Antidot1,Antidot2-1,Antidot2-2,Antidot3-1,Antidot3-2,Antidot4,Antidot5}%
. This artificial crystals lead to a wide variety of phenomena, for instance,
Weiss oscillation, chaotic dynamics of electrons, the formation of an
electronic miniband structure and massless Dirac fermions. At low
temperatures, the mean free path of electrons is much longer than the period
of antidot lattices ranging from 10 to 100 nanometers. The modulated periodic
potential can also be created by electron beam lithography electrodeposition
and periodic arrays of metallic nanodots can be realized on semiconductor
surfaces. Due to elastic strains producing these dots, a sufficiently strong
piezoelectric potential modulation results in miniband effects in the
underlying 2DEG \cite{Antidot2-1,Antidot2-2}. Very recently, a honeycomb
lattice of coronene molecules was\ created by using a cryogenic scanning
tunneling microscope on a Cu(111) surface to construct artificial
graphene-like lattice with the lattice constant approaching $5$nm
\cite{Antidot5}.

\begin{figure}[tbh]
\includegraphics[width=1.0\columnwidth]{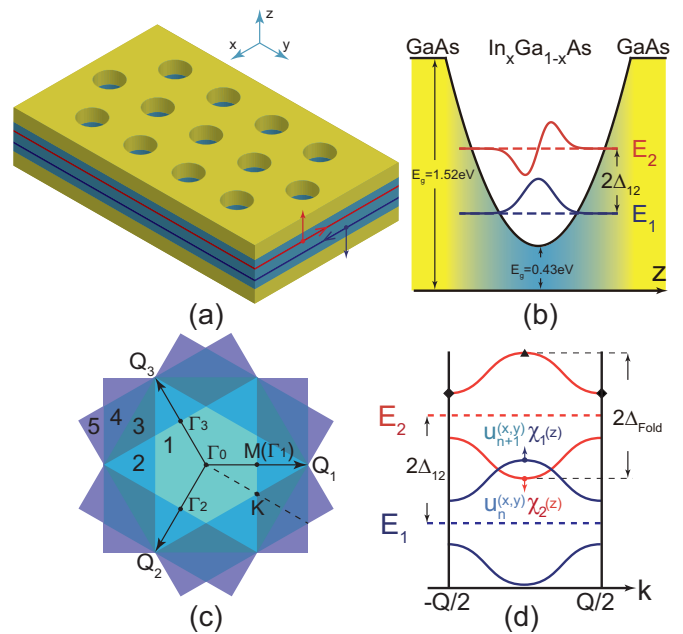}\caption{(color online)
Schematic of the proposed structure and its energy bands. (a) A GaAs/In$_{x}%
$Ga$_{1-x}$As/GaAs parabolically graded QW with an antidot lattice, which can
be created by etching technique. (b) Band profile and the first and second
subbands of the parabolically graded QW. (c) Brillouin zone folding induced by
a triangular antidot lattice. The numbers $1-5$ denote the first to the fifth
Brillouin zones of the antidot lattice. (d) Minibands of the antidot lattice
from folding the first and second subbands of the QW ($Q=2\pi/a$ and $a$ is
the antidot lattice constant). Note that band inversion occurs between
neighboring minibands.}%
\label{fig1}%
\end{figure}

We consider the 2DEG in a GaAs/In$_{x}$Ga$_{1-x}$As/GaAs PQW, which was
fabricated successfully before \cite{paraQW1-1,paraQW1-2,paraQW2}, with a
triangular antidot lattice (see Fig.~1(a)). Before going to the numerical
calculation, we first give a clear physical picture for the emergence of a TI
phase in this 2DEG system upon nanostructuring with antidot lattice. The
simplest description of the 2DEG system is obtained by reducing the eight-band
Kane model to the lowest conduction subbands of the QW (see Appendix B). This
gives the Hamiltonian for the 2DEG with periodic antidot lattice potential
$V(x,y)$:%

\begin{equation}
H=\hbar^{2}k^{2}/2m+\Delta_{12}\tau_{z}+\eta\tau_{x}(k_{x}\sigma^{x}%
+k_{y}\sigma^{y})+V(x,y),
\end{equation}
where $\tau_{i}$ $(i=x,y,z)$ are Pauli matrices describing the first and
second QW subbands $\{\chi_{n}(z)\}$ $(n=1,2)$ of effective mass $m$, and
$\sigma^{i}$ $(i=x,y,z)$ refer to the electron spin. The second term
$\Delta_{12}\tau_{z}$ comes from the energy difference $2\Delta_{12}$ between
the first and second subbands [see Fig. 1(b)]. The third term $\eta\tau
_{x}(k_{x}\sigma^{x}+k_{y}\sigma^{y})$ describes the inter-subbands SOI (ISOI)
obtained from the eight-band Kane model using the L\"{o}wdin perturbation
theory (see Appendix B). The coupling strength $\eta$ is
\begin{equation}
\eta=\frac{1}{3}\left\langle \chi_{2}(z)\right\vert \sum_{i=g,g^{\prime}%
}\left(  P^{2}(z)\frac{\partial_{z}E_{i}(z)}{E_{i}^{2}(z)}+P(z)\frac
{\partial_{z}P(z)}{E_{i}(z)}\right)  \left\vert \chi_{1}(z)\right\rangle ,
\label{YITA}%
\end{equation}
where $E_{g}(z)$ and $\Delta_{0}(z)$ are the band gap and spin split-off
splitting in the QW region, $E_{g^{\prime}}(z)\equiv E_{g}(z)+\Delta_{0}(z)$,
and $P(z)$ is the Kane matrix element. The SOIs in 2DEGs usually come from the
asymmetry of the QWs, i.e., Rashba SOI. Surprisingly, the ISOI can appear in a
symmetric PQW, behaving like a hidden SOI. From Eq. (\ref{YITA}), one can see
that the ISOI arises from the spatial variations of the bandgap $E_{g}(z)$,
the Kane matrix $P(z)$, and the intrinsic SOI $\Delta_{0}(z)$, i.e., the
variation of the concentration of \textit{In} component, which behaves like an
effective local electric field. This local electric field would not push the
electron and the hole states to the left and right sides of the QW, but it can
induce a considerably large ISOI hidden in symmetric QWs. The initial
$\left\vert \chi_{1}(z)\right\rangle $ and final states $\left\vert \chi
_{2}(z)\right\rangle $ are neighboring subbands having opposite parity, while
the variations of $E_{g}(z)$, $P(z)$ and $\Delta_{0}(z)$ in a symmetric QW are
odd. This means that the ISOI can exist in symmetric QWs. \begin{figure}[tbh]
\includegraphics[width=1.0\columnwidth]{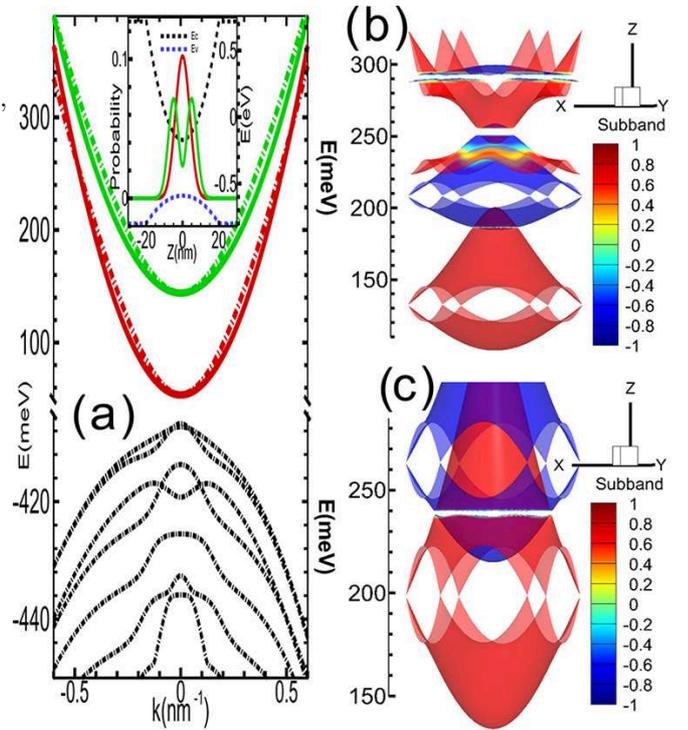}\caption{(color online) (a)
Band structure of a GaAs/In$_{x}$Ga$_{1-x}$As/GaAs parabolically graded QW
from eight-band Kane model. The red and blue dashed (solid) curves denote the
first and second subbands from the eight-band Kane model (the reduced
four-band model). The inset shows the spatial distributions of the first and
second subbands. (b) and (c): minibands of a parabolically graded QW for two
different triangular antidot lattices: the lattice constants and barrier
heights $V_{0}$ of the antidot lattices are $a=17$ nm, $V_{0}=200$ meV for (b)
and $12.5$ nm, $300$ meV for (c). The minibands are inverted and the minigaps
are opened.}%
\end{figure}

The first and second subbands both form minibands due to the Brillouin zone
folding caused by the antidot lattice. The band inversion could occur between
the two adjacent minibands of the first subband $\psi_{+}=\chi_{1}%
(z)u_{k,n_{1}}(x,y)$ and the second subband $\psi_{-}=\chi_{2}(z)u_{k,n_{2}%
}(x,y)$ (see Fig. 1(d)), where $u_{k,n}(x,y)$ is the $n$th miniband formed by
the antidot lattice. We model the triangular antidot lattice potential
$V\left(  x,y\right)  $ by a periodic potential $V_{0}[1+%
%TCIMACRO{\tsum \nolimits_{i=1}^{3}}%
%BeginExpansion
{\textstyle\sum\nolimits_{i=1}^{3}}
%EndExpansion
\cos\left(  \mathbf{Q}_{i}\cdot\mathbf{r}\right)  ]$ with potential height
$V_{0}$, $\mathbf{Q}_{1}=(2\pi/a)(1,0)$, $\mathbf{Q}_{2}=(2\pi/a)(-1/2,-\sqrt
{3}/2)$, $\mathbf{Q}_{3}=(2\pi/a)(-1/2,\sqrt{3}/2)$ and $a$ is the triangular
antidot lattice constant (see Appendix C).

To describe the four minibands (two spin-degenerate minibands) $\left\vert
\psi_{+},\uparrow\right\rangle $, $\left\vert \psi_{-},\downarrow\right\rangle
$, $\left\vert \psi_{+},\downarrow\right\rangle $, $\left\vert \psi
_{-},\uparrow\right\rangle $ involved in the band inversion, we treat other
electron and hole minibands by L\"{o}wdin perturbation theory and reduce the
eight-band Kane \textbf{k}$\cdot$\textbf{p} model to the following effective
Hamiltonian within the basis ($\left\vert \psi_{+},\uparrow\right\rangle $,
$\left\vert \psi_{-},\downarrow\right\rangle $, $\left\vert \psi
_{+},\downarrow\right\rangle $, $\left\vert \psi_{-},\uparrow\right\rangle $):%
\begin{equation}
H_{\mathrm{eff}}^{4\times4}=\left(
\begin{array}
[c]{cccc}%
M-Bk^{2} & Ak_{-} & 0 & 0\\
Ak_{+} & -M+Bk^{2} & 0 & 0\\
0 & 0 & M-Bk^{2} & Ak_{+}\\
0 & 0 & Ak_{-} & -M+Bk^{2}%
\end{array}
\right)  , \label{BHZ}%
\end{equation}
(see Appendix D and E), which assumes the same form as the $D=0$ BHZ
Hamiltonian. Here $k_{\pm}=k_{x}\pm ik_{y}$, $M=\Delta_{12}-\Delta
_{\mathrm{Fold}}$ [see Fig. 1(b)], $B=-\hbar^{2}/2m^{\ast}$, and $A$
characterize the ISOI strength between neighboring minibands with opposite
spin. This Hamiltonian obviously has a $Z_{2}$ topological phase when $M<0$,
corresponding to band inversion. \begin{figure}[tbh]
\includegraphics[width=1.0\columnwidth]{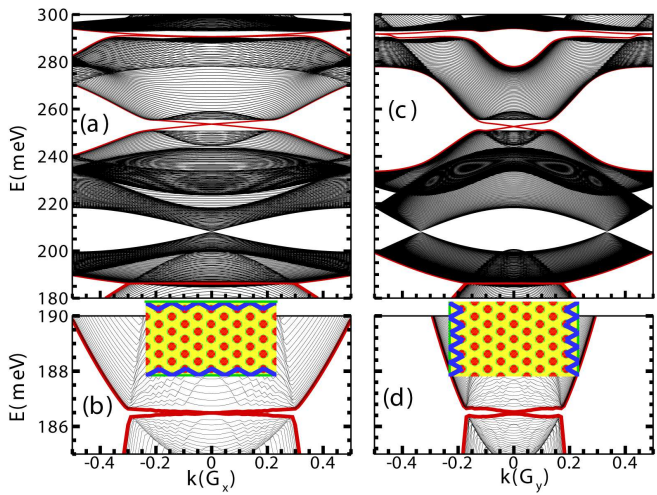}\caption{(color online) Edge
states in nontrivial minigaps of a Hall bar structure of the antidot lattice.
The Hall bar orientation is along the \textit{x} axis (a) or \textit{y} axis
(c), as sketched in the insets of (b) and (d), respectively. The lower
panels\textit{ }(b) and (d) amplify the lowest nontrivial minigaps and gapless
topological edge states. The lattice constant of the antidot lattice is $a=17$
nm. The blue curves indicate the spatial distributions of the edge states.}%
\end{figure}

Next we employ the eight-band Kane \textbf{k}$\cdot$\textbf{p} model to
calculate the subband structure with SOIs in a 40-nm-thick GaAs/In$_{x}%
$Ga$_{1-x}$As/GaAs PQWs \cite{paraQW1-1,paraQW1-2}, as plotted in Fig. 2(a).
The energy difference between the minima of the first and second subbands at
$\Gamma$ point is about 90 meV [see Fig. 2(a)]. In order to calculate the
miniband structures caused by an in-plane periodic potential induced by the
triangular antidot lattice, we reduce the eight-band model to an effective
four-band \textbf{k}$\cdot$\textbf{p} Hamiltonian by including the lowest 20
electron subbands and 54 highest hole subbands in the QW, to reproduce the
energy dispersions of the first and second subbands calculated from the
eight-band Kane model [see Fig. 2(a)]. The parameters in the four-band
Hamiltonian is given in Appendix E. The minibands from the four-band
\textbf{k}$\cdot$\textbf{p} Hamiltonian are shown in Figs. 2(b) and 2(c).
These minibands originates from folding the first and second subbands of the
QW into the first Brillouin zone of the antidot lattice [Fig. 1(c)]. By tuning
the antidot lattice constant $a$ and the potential height $V_{0}$, i.e., the
etching depth of the antidot lattice, many band inversions appear between
these minibands, which can be clearly seen in Figs. 2(b) and 2(c). The
minigaps between these minibands are opened by the ISOI shown in Eq.
(\ref{YITA}) [see Figs. 2(b) and 2(c)].

To demonstrate that these minigaps are topologically nontrivial, we determine
the parity of each miniband at the four time-reversal invariant momenta
\cite{Fu2} $\Gamma_{i}$ ($i=0,1,2,3$) in the first Brillouin zone shown in
Fig. 1(c). For the lowest $N$ spin-degenerate minibands being occupied, the
$Z_{2}$ invariant is given by $\left(  -1\right)  ^{\nu}=%
%TCIMACRO{\tprod \nolimits_{i}}%
%BeginExpansion
{\textstyle\prod\nolimits_{i}}
%EndExpansion
\delta_{i},\delta_{i}=%
%TCIMACRO{\tprod \nolimits_{m=1}^{N}}%
%BeginExpansion
{\textstyle\prod\nolimits_{m=1}^{N}}
%EndExpansion
\xi_{2m}\left(  \Gamma_{i}\right)  $, where $\xi_{2m}\left(  \Gamma
_{i}\right)  =\pm1$ is the parity of the $2m$th occupied miniband at
$\Gamma_{i}$. Our calculation gives $\nu=1$ at all the minigaps, which proves
the whole system is in the quantum spin-Hall phase (see Appendix F).

Next, we demonstrate the emergence of topological edge states upon etching the
QW into a Hall bar structure along two different directions ($x$ axis and $y$
axis). As shown in Fig. 3, a pair of topological helical edge states appear
inside each nontrivial minigap. For example, we can see topological helical
edge states in the lowest two nontrivial minigaps near $\sim186.5$ meV and
$\sim255$ meV, respectively. The helical edge state pairs in these minigaps
would lead to higher conductance plateaus as the Fermi energy increases by
increasing the doping level. The helical edge states do not overlap with the
bulk states, making it possible to be detected experimentally.

The lowest nontrivial minigaps is quite small (about $0.5$ meV), but the
second minigap is larger (about $5$ meV). By tuning the period and potential
height of the antidot lattice, the nontrivial minigaps can be significantly
enhanced [see Figs. 4(a) and 4(b)]. For example, the lowest minigaps can be
enhanced to $5$ meV, which is already comparable with that in HgTe and
InAs/GaSb QW systems ($\sim10$ meV) \cite{Konig1,Konig2}. The second minigaps
can approach $20$ meV, which means the TI phase can be realized at liquid
nitrigen temperature regime. From Figs. 4(a) and 4(b), one can see that the
lowest nontrivial minigap is closed as the lattice constant $a$ increases, but
the second higher nontrivial minigap survives, i.e., the TI phase can exist
even at large lattice constants, e.g., $25$ nm.

Finally, we comment on the experimental detection the aforementioned edge
states in GaAs/In$_{x}$Ga$_{1-x}$As/GaAs quantum Hall bar (shown schematically
in the insets of Figs. 3(b) and 3(d)). One way to detect the aforementioned
edge states (shown in Fig. 3) is the standard four terminal measurements as
demonstrated in previous works \cite{Konig1,Konig2}. In contrast to HgTe and
InAs/GaSb QW systems, there are many pairs of helical edge states in our
system between these inverted minibands, which leads to higher plateaus with
increasing the Fermi energy. Another possible way is microwave impedance
microscopy which makes spatial-resolved nano-scale images (%
%TCIMACRO{\TEXTsymbol{<} }%
%BeginExpansion
$<$
%EndExpansion
100 nm) of the conductivity and permittivity of a sample \cite{MIM}. The
unoccupied edge states in higher minigaps can be detected using the
angle-resolved photonemission technique \cite{ARPAS2013}, which has already
been successfully applied to identify occupied and unoccupied surface states
in Bi$_{2}$Se$_{3}$ and Bi$_{2}$Te$_{x}$Se$_{3}$
\cite{ARPAS1999,ARPAS2012,ARPAS2013}.

\begin{figure}[tbh]
\includegraphics[width=1.0\columnwidth]{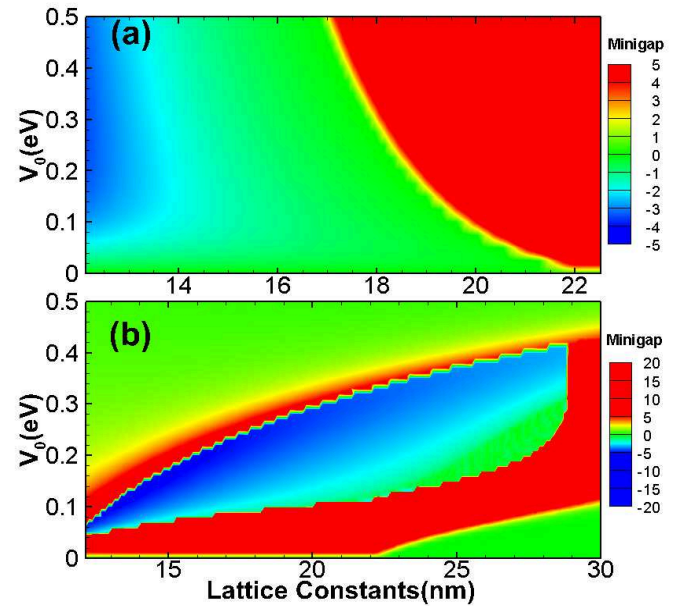}\caption{Phase diagrams of
the antidot lattice on a GaAs/In$_{x}$Ga$_{1-x}$As/GaAs parabolically graded
QWs. The lowest (a)\ and second (b) minigap vs. antidot lattice constant $a$
and potential height $V_{0}$. Negative (Positive)\ minigap indicates TI phase
(normal phase).}%
\end{figure}

\section{CONCLUSION}

Our proposal is based on a general analysis about the electron orbital motion
in TIs. By using the Born-Oppenheimer approximation, we find that the fast
motion will induce a spin-dependent gauge field on slow orbital motion. Based
on this general analysis, we demonstrate theoretically the TI phase in a
conventional 2DEG embedded in a symmetric GaAs/In$_{x}$Ga$_{1-x}$As/GaAs PQW,
with antidot lattices created by well-developed etching technique. The key
point is to create a ISOI in a symmetric QW, in contrast to conventional SOI
in asymmetric QWs. This hidden ISOI in symmetric QWs induces a spin-dependent
effective Lorentz force on the electrons, and generates the TI phases in such
system. Interestingly, such ISOI exists in conventional semiconductors with a
positive bandgap, i.e., normal band structures can generate quite large
nontrivial gaps approaching 20 meV. This make it possible to observe the
quantum spin Hall effect in liquid nitrigen temperature regime.

So far, all members of TI family are narrow bandgap systems containing heavy
atoms. Our proposal breaks this constraint, and makes it possible to realize
TI phase in conventional semiconductor 2DEG using the well-developed
semiconductor fabrication techniques \cite{Kono,Giti,Heremans}. The presence
of the TI phase in PQWs with antidot lattice can largely advance the
application of this new quantum state in existing electronics and
optoelectronics devices. The general designing principle proposed in this
work, i.e., the gauge field acting on slow orbital motion induced by interband
coupling, paves a new way for generating nontrivial topological phases, such
as quantum spin Hall phase and even quantum anomalous Hall phases by doping
magnetic ions, in conventional semiconductor 2DEGs, and suggests a promising
approach to integrate it in well developed semiconductor electronic devices.

\section{ACKNOWLEDGMENTS}

This work was supported by the NSFC Grants Nos. 11434010, 11304306 and the
grant No. 2011CB922204 from the MOST of China. KC\ would like to appreciate
Prof. S. C. Zhang for helpful discussions. LKS and WKL contributed equally to
this work.

\section{APPENDIX A: EFFECTIVE GAUGE FIELD IN SPIN-ORBIT COUPLED SYSTEMS}

In the spin-orbit coupled system, adopting the Born-Oppenheimer approximation,
the\ total Hamiltonian can be divided into two parts:%
\begin{equation}
\hat{H}(\mathbf{\hat{k}},\mathbf{\hat{r}},\boldsymbol{\hat{\sigma}})=\hat
{H}_{\mathrm{orb}}(\mathbf{\hat{k}},\mathbf{\hat{r}})+H_{\mathrm{s-o}%
}(\mathbf{\hat{k}},\boldsymbol{\hat{\sigma}}),\nonumber
\end{equation}
where $\hat{H}_{\mathrm{orb}}(\mathbf{\hat{k}},\mathbf{\hat{r}})$ stands for
the intra-band (slow) orbital motion part, and $H_{\mathrm{s-o}}%
(\mathbf{\hat{k}},\boldsymbol{\hat{\sigma}})$ is the inter-band (fast)
spin-orbit part. For a given eigenvalue $\mathbf{k}$ of the momentum operator
$\mathbf{\hat{k}}$, the eigenstates of the spin-orbit part is denoted by
$|n(\mathbf{k})\rangle$ ($n=1,2,\cdots$) and the corresponding eigenenergies
are $\epsilon_{n}(\mathbf{k})$. We work in the momentum representation of the
orbital part and expand the eigenstate of $\hat{H}(\mathbf{\hat{k}%
},\mathbf{\hat{r}},\boldsymbol{\hat{\sigma}})$ in this representation,
$|\Psi(\mathbf{k})\rangle\equiv\langle\mathbf{k}|\Psi\rangle$, as%
\[
|\Psi(\mathbf{k})\rangle=\sum_{n,\mathbf{k}}\phi_{n}(\mathbf{k})|\chi
_{n}(\mathbf{k})\rangle.
\]
In the momentum representation, we have $\mathbf{\hat{r}}=i\nabla_{\mathbf{k}%
}$ and $\mathbf{\hat{k}}=\mathbf{k}$. Substituting into $\hat{H}%
(\mathbf{k},\mathbf{\hat{r}},\boldsymbol{\hat{\sigma}})|\Psi(\mathbf{k}%
)\rangle=E|\Psi(\mathbf{k})\rangle$, we have%
\[
\sum_{n}H_{m,n}(\mathbf{k})\phi_{n}(\mathbf{k})=E\phi_{m}(\mathbf{k}),
\]
where
\begin{align*}
H_{m,n}(\mathbf{k})  &  \equiv\delta_{m,n}\epsilon_{m}(\mathbf{k})+\langle
\chi_{m}(\mathbf{k})|\hat{H}_{\mathrm{orb}}(\mathbf{k},\mathbf{\hat{r}}%
)|\chi_{n}(\mathbf{k})\rangle\\
&  =\delta_{m,n}\epsilon_{m}(\mathbf{k})+\hat{H}_{\mathrm{orb}}(\mathbf{k}%
,\mathbf{\hat{r}}-\mathbf{A}_{m,n}(\mathbf{k}))
\end{align*}
contains a pure gauge $\mathbf{A}_{m,n}(\mathbf{k})\equiv-i\langle\chi
_{m}(\mathbf{k})|\nabla_{\mathbf{k}}|\chi_{n}(\mathbf{k})\rangle$. By now the
above equation is still exact. Now we make the Born-Oppenheimer approximation
and consider adiabatic transport, i.e., neglect the off-diagonal coupling
between different spin-orbit energy bands, to arrive at the single-band
description
\[
H_{n}(\mathbf{k},\mathbf{\hat{r}}_{n}^{\prime})\phi_{n}(\mathbf{k})=E\phi
_{n}(\mathbf{k}),
\]
where the effective single-band Hamiltonian on the orbital motion
\[
H_{n}(\mathbf{k},\mathbf{\hat{r}}_{n}^{\prime})=\epsilon_{n}(\mathbf{k}%
)+\hat{H}_{\mathrm{orb}}(\mathbf{k},\mathbf{\hat{r}}_{n}^{\prime}),
\]
where $\mathbf{\hat{r}}_{n}^{\prime}=\mathbf{\hat{r}}_{n}-\mathbf{A}%
_{n}(\mathbf{k})$ contains an effective gauge field for the slow orbital
motion:
\[
\mathbf{A}_{n}(\mathbf{k})=-i\left\langle \chi_{n}(\mathbf{k})\right\vert
\nabla_{\mathbf{k}}\left\vert \chi_{n}(\mathbf{k})\right\rangle .
\]

Specifically, for the BHZ model of 2D TIs in the presence of an in-plane
uniform electric field, the slow orbital part is%
\begin{equation}
\hat{H}_{\mathrm{orb}}(\mathbf{\hat{k}},\mathbf{\hat{r}})=C-Dk^{2}%
-e\mathbf{E}\cdot\mathbf{r},\nonumber
\end{equation}
and the fast spin-orbital part $H_{\mathrm{s-o}}(\mathbf{\hat{k}%
},\boldsymbol{\hat{\sigma}})$ is
\begin{equation}
\left(
\begin{array}
[c]{cccc}%
M-Bk^{2} & Ak_{+} & 0 & 0\\
Ak_{-} & -M+Bk^{2} & 0 & 0\\
0 & 0 & M-Bk^{2} & Ak_{-}\\
0 & 0 & Ak_{+} & -M+Bk^{2}%
\end{array}
\right)  ,\nonumber
\end{equation}
and%
\begin{equation}
A_{x,n}=s_{n}\frac{k_{y}}{2k^{2}}\left[  1+t_{n}\frac{Bk^{2}-M}{[A^{2}%
k^{2}+(M-Bk^{2})^{2}]^{1/2}}\right]  ,\nonumber
\end{equation}%
\begin{equation}
A_{y,n}=s_{n}\frac{k_{x}}{2k^{2}}\left[  1+t_{n}\frac{Bk^{2}-M}{[A^{2}%
k^{2}+(M-Bk^{2})^{2}]^{1/2}}\right]  .\nonumber
\end{equation}
where $s_{n}=\pm1$ for spin-up/down block, and $t_{n}=\pm1$ for the electron
in conduction/valence band ($n=1,2,3,4$). The effective vector potential leads
to the non-trivial effective gauge field with the strength%
\begin{align}
F_{xy,n}(\mathbf{k})  &  \equiv i[x^{\prime},y^{\prime}]=i[i\partial_{k_{x}%
}-A_{x},i\partial_{k_{y}}-A_{y}]\ \nonumber\\
&  =(\nabla_{\mathbf{k}}\times\mathbf{A})_{z}=\lambda_{n}\frac{A^{2}%
(M+Bk^{2})}{2[A^{2}k^{2}+(M-Bk^{2})^{2}]^{3/2}},\nonumber
\end{align}
where $\lambda_{n}=s_{n}\times t_{n}=\pm1$ ($n=1,2,3,4$). Within the
Born-Oppenheimer approximation, the equation of motion for the $n$-th band can
be written as
\begin{align}
{\dot{x}}_{n}^{\prime}  &  =\frac{\partial H_{n}}{\hbar\partial k_{x}%
}+F_{xy,n}(k){\dot{k}}_{y},\nonumber\\
{\dot{y}}_{n}^{\prime}  &  =\frac{\partial H_{n}}{\hbar\partial k_{x}%
}-F_{xy,n}(k){\dot{k}}_{x},\nonumber\\
{\dot{k}}_{i}  &  =eE_{i}/\hbar,\nonumber
\end{align}
we can see that the gauge field strength $F_{xy,n}=(\nabla_{\mathbf{k}}%
\times\mathbf{A})_{z}$ acts as a Lorentz force in the $k$-space, acting on
spin-up and spin-down electrons in opposite directions, which is perpendicular
to electron momentum.

\section{APPENDIX B: EFFECTIVE SPIN-ORBIT COUPLING IN A QUANTUM WELL}

For a symmetric QW grown along (001) direction (the \textit{z} axis),
effective spin-orbit coupling exists between subbands with opposite parities.
This effective spin-orbit coupling comes from interband coupling and can be
understand by reducing the 8$\times$8 Kane Hamiltonian to a 2$\times$2
effective Hamiltonian \cite{Lowdin}.

To the first order of $k$, the 8$\times$8 Kane Hamiltonian in the basis
($i\left\vert S,\uparrow\right\rangle $, $i\left\vert S,\downarrow
\right\rangle $, $\left\vert 3/2,1/2\right\rangle $, $\left\vert
3/2,-1/2\right\rangle $, $\left\vert 3/2,3/2\right\rangle $, $\left\vert
3/2,-3/2\right\rangle $, $\left\vert 1/2,1/2\right\rangle $, $\left\vert
1/2,-1/2\right\rangle $) around the $\Gamma$ point is%
\[
H_{8\times8}=\left(
\begin{array}
[c]{cc}%
H_{c} & H_{cv}\\
H_{cv}^{\dag} & H_{v}%
\end{array}
\right)  ,
\]
where $H_{c}=$ $\epsilon_{c}I_{2\times2}$ and $H_{v}=\epsilon_{v}I_{4\times
4}\oplus\epsilon_{s}I_{2\times2}$ are 2$\times$2 and 6$\times$6 diagonal part
for conduction and valence bands, and the 2$\times$6 matrix%
\[
H_{cv}=\left(
\begin{array}
[c]{cccccc}%
\frac{-\sqrt{2}Pk_{z}}{\sqrt{6}} & \frac{-Pk_{-}}{\sqrt{6}} & \frac{Pk_{+}%
}{\sqrt{2}} & 0 & \frac{Pk_{z}}{\sqrt{3}} & \frac{-Pk_{-}}{\sqrt{3}}\\
\frac{Pk_{+}}{\sqrt{6}} & -\frac{\sqrt{2}Pk_{z}}{\sqrt{3}} & 0 & \frac{Pk_{-}%
}{\sqrt{2}} & \frac{Pk_{+}}{\sqrt{3}} & \frac{Pk_{z}}{\sqrt{3}}%
\end{array}
\right)
\]
represents the interband coupling. Specifically, $\epsilon_{i}=\hbar^{2}%
k^{2}/(2m)+V_{i}$ is the kinetic energy plus the total potential for the
conduction/valence/spin-split ($i=c/v/s$) bands, with $V_{c}-V_{v}=E_{g}$ the
band gap and $V_{v}-V_{s}=\Delta_{0}$ the band off set. $k_{\pm}=k_{x}\pm
ik_{y}$ and $P=-i(\hbar/m_{0})\left\langle S_{c}\right\vert p_{x}\left\vert
X_{v}\right\rangle $ parameterize the interband coupling.

The eigenvalue problem can be expressed as%
\[
\left(
\begin{array}
[c]{cc}%
H_{c} & H_{cv}\\
H_{cv}^{\dag} & H_{v}%
\end{array}
\right)  \left(
\begin{array}
[c]{c}%
\varphi_{c}\\
\varphi_{v}%
\end{array}
\right)  =\varepsilon\left(
\begin{array}
[c]{c}%
\varphi_{c}\\
\varphi_{v}%
\end{array}
\right)  ,
\]
where $\varphi_{c}$ is a two-component spinor for conduction bands and
$\varphi_{v}$ is a six-component spinor for valence bands. Since we focus on
the conduction bands, $\varphi_{v}=\left(  \varepsilon-H_{v}\right)
^{-1}H_{cv}^{\dag}\psi_{c}$ can be eliminated and gives the effective
Schr\"{o}dinger-type equation $H_{\mathrm{eff}}\varphi_{c}=\varepsilon
\varphi_{c}$, with $H_{\mathrm{eff}}=H_{c}+H_{cv}\left(  \varepsilon
-H_{v}\right)  ^{-1}H_{cv}^{\dag}$ for conduction bands. Without loss of
generality, we assume the QW is non-uniform only along the $z$ direction,
e.g., a PQW. By straightforward algebra, we have $H_{\mathrm{eff}}%
^{\uparrow\downarrow}=\left(  H_{\mathrm{eff}}^{\downarrow\uparrow}\right)
^{\dag}=\sum_{i=v,s}(k_{-}/3)P(z)\left[  (\varepsilon-\epsilon_{i}%
(z))^{-1}P(z),k_{z}\right]  $, where $H_{\mathrm{eff}}^{\uparrow\downarrow}$
and $H_{\mathrm{eff}}^{\downarrow\uparrow}$ represent the effective spin-orbit
coupling between the spin up and down electron.

Since we focus on the lowest conduction subbands, we have $\varepsilon
-\epsilon_{v}(z)\approx E_{g}(z)$ and $\varepsilon-\epsilon_{s}(z)\approx
E_{g}(z)+\Delta_{0}(z)\equiv E_{g^{\prime}}(z)$. Because $E_{g}(z)$ and
$E_{g^{\prime}}(z)$ are much larger than the subband energies in the wide QWs
under consideration, we keep the zero-th order terms $E_{g}^{-1}(z)$ and
$E_{g^{\prime}}(z)^{-1}$ in the expansion, and project the spin-orbit coupling
operator $H_{\mathrm{eff}}^{\uparrow\downarrow}$ into the two lowest
spin-degenerate subbands ($\left\vert \chi_{1}(z)\right\rangle $, $\left\vert
\chi_{2}(z)\right\rangle $) to obtain the ISOI $\eta\tau_{x}(k_{x}\sigma
^{x}+k_{y}\sigma^{y})$, where $\eta$ is given in Eq. (\ref{YITA}), $\sigma
^{i}$ denotes the real electron spin, and $\tau_{i}$ refers to the Pauli
matrix describing the the subband index.

\section{APPENDIX C: BAND EDGE WAVE FUNCTIONS IN FOLDED BRILLOUIN ZONE}

We consider a PQW in the presence of an antidot lattice, which can be
generally described by a potential $V(\mathbf{r})=\sum_{\mathbf{q}}\tilde
{V}(\mathbf{q})e^{i\mathbf{q}\cdot\mathbf{r}}$ with the lattice periodicity.

For a triangular antidot lattice, the reciprocal lattice vectors in the
hexagonal Brillouin zone are $\mathbf{Q}_{1}=\left(  2\pi/3a\right)  (3,0)$,
$\mathbf{Q}_{2}=\left(  2\pi/3a\right)  (-3/2,-3\sqrt{3}/2)$, $\mathbf{Q}%
_{3}=-(\mathbf{Q}_{1}+\mathbf{Q}_{2})$. The envelope functions of the lowest
miniband at the band edge ($k=0$, $\Gamma$ point) is $u_{\Gamma,1}%
(\mathbf{r})=1-%
%TCIMACRO{\tsum \nolimits_{l=1}^{3}}%
%BeginExpansion
{\textstyle\sum\nolimits_{l=1}^{3}}
%EndExpansion
\left[  2m\tilde{V}(\mathbf{Q}_{l})/\hbar^{2}Q^{2}\right]  \cos(\mathbf{Q}%
_{l}\cdot\mathbf{r})$. For higher minibands, their envelope functions
$u_{\Gamma,n}(\mathbf{r})$ ($n=2,3,4,5,6,7$) at the band edge are linear
combinations of the six wave vector components ($e^{\pm i\mathbf{Q}_{1}%
\cdot\mathbf{r}}$, $e^{\pm i\mathbf{Q}_{2}\cdot\mathbf{r}}$, $e^{\pm
i\mathbf{Q}_{3}\cdot\mathbf{r}}$), e.g., $u_{\Gamma,2}(\mathbf{r})\propto
\sum_{l=1}^{3}\sin(\mathbf{Q}_{l}\cdot\mathbf{r})$ and $u_{\Gamma
,7}(\mathbf{r})\propto\sum_{l=1}^{3}\cos(\mathbf{Q}_{l}\cdot\mathbf{r})$. The
most important minibands are $u_{k,1}(r)$, $u_{k,2}(r)$ and $u_{k,4}(r)$: the
lowest nontrivial minigap occurs between $u_{k,1}(r)$ and $u_{k,2}(r)$, and
the second nontrivial minigap occurs between $u_{k,2}(r)$ and $u_{k,4}(r)$.

\section{APPENDIX D: EFFECTIVE BHZ HAMILTONIAN NEAR THE $\Gamma$ POINT}

The lowest two subbands $\chi_{1}(z)$ and $\chi_{2}(z)$ in a PQW have even and
odd opposite parities, an effective spin-orbit interaction $\left\langle
\chi_{1}(z)\right\vert \eta(z)(k_{x}\sigma^{x}+k_{y}\sigma^{y})\left\vert
\chi_{2}(z)\right\rangle $ appears. When the Brillouin zone is folded by the
triangular anti-dot lattice, the lowest nontrivial minigap appears between the
miniband pair $\left\vert \chi_{1}(z),u_{k,2}(r)\right\rangle $ and
$\left\vert \chi_{2}(z),u_{k,1}(r)\right\rangle $, i..e., the second miniband
of the first subband and the first miniband of the second subband. The second
nontrivial minigap appears between the miniband pair $\left\vert \chi
_{2}(z),u_{k,2}(r)\right\rangle $ and $\left\vert \chi_{1}(z),u_{k,4}%
(r)\right\rangle $, i.e., the second miniband of the second subband and the
fourth miniband of the first subband. To obtain an effective Hamiltonian near
each minigap, we project the Hamiltonian $H=\hbar^{2}k^{2}/2m+\Delta_{12}%
\tau_{z}+\eta\tau_{x}(k_{x}\sigma^{x}+k_{y}\sigma^{y})+V(x,y)$ onto the
corresponding miniband pair and obtain an effective BHZ model Eq. (\ref{BHZ})
in the basis $\left\vert \psi_{+},\uparrow\right\rangle $, $\left\vert
\psi_{-},\downarrow\right\rangle $, $\left\vert \psi_{+},\downarrow
\right\rangle $, $\left\vert \psi_{-},\uparrow\right\rangle $, where
$|\psi_{+}\rangle$ is the miniband above $\left\vert \psi_{-}\right\rangle $
by\ $2M$ at the $\Gamma$ point, $B=-\hbar^{2}/2m^{\ast}$ characters the band
dispersions with the effective mass $m^{\ast}$ near the band edge, and $A$
characterize the intersubband spin-orbit coupling. At the $\Gamma$ point%
\begin{align*}
&  \left\langle \psi_{+}\right\vert \eta(z)(k_{x}\sigma^{x}+k_{y}\sigma
^{y})\left\vert \psi_{-}\right\rangle \\
&  =\left\langle \chi_{1(2)}(z)\right\vert \eta(z)\left\vert \chi
_{2(1)}(z)\right\rangle \\
&  \cdot\left\langle u_{k,2(2)}(r)\right\vert (-i\partial_{x}\sigma
_{x}-i\partial_{y}\sigma_{y})\left\vert u_{k,1(4)}(r)\right\rangle \\
&  =A\left(  k_{x}\sigma^{x}+k_{y}\sigma^{y}\right)  .
\end{align*}
The accurate coupling strength can be estimated by numerical calculating based
on the eight-band Kane model.

For BHZ model, a $Z_{2}$ topological transition from the normal phase to the
topological insulator phase would occur when $M=\Delta_{12}-\Delta
_{\text{\textrm{Fold}}}$ [see Fig. 1(b)] changes sign from positive to
negative, which can be controlled by adjusting the lattice constants and
etching depths of antidots.

\section{APPENDIX E: THE EFFECTIVE HAMILTONIAN REDUCED NUMERICALLY FROM THE
EIGHT-BAND KANE MODEL}

At the $\Gamma$ point, the wave functions in the eight-band
$\boldsymbol{k\cdot p}$ Hamiltonian are%
\begin{equation}
\psi_{m}=\left(
\begin{array}
[c]{c}%
F_{1}^{(m)}\left(  z\right) \\
F_{2}^{(m)}\left(  z\right) \\
F_{3}^{(m)}\left(  z\right) \\
F_{4}^{(m)}\left(  z\right) \\
\vdots\\
F_{8}^{(m)}\left(  z\right)
\end{array}
\right)  e^{i\mathbf{k}_{\parallel}\cdot\mathbf{\vec{r}}},\nonumber
\end{equation}
which can be obtained by solving the secular equation $H_{8\times8}^{\left(
0\right)  }\psi_{m}=E_{m}\psi_{m}$.

Considering the two lowest electron subbands, we obtain the effective
two-dimensional Hamiltonian by averaging the $z$ component in the Hamiltonian%
\begin{equation}
H_{\mathrm{eff}}(\mathbf{k}_{\parallel})=\langle\Psi(z)|H|\Psi(z)\rangle
,\nonumber
\end{equation}
where the matrix element of the Hamiltonian is
\begin{align}
\left\langle H_{\mathrm{eff}}\right\rangle _{mn}  &  =\left\langle
\psi^{\left(  m\right)  }\right\vert H_{8\times8}\left\vert \psi^{\left(
n\right)  }\right\rangle \nonumber\\
&  =\sum_{i,j=1}^{8}\langle F_{i}^{\left(  m\right)  }\left(  z\right)
|H_{ij}|F_{j}^{\left(  n\right)  }\left(  z\right)  \rangle.\nonumber
\end{align}

The Hamiltonian can be divided into
\begin{align}
H  &  =H^{\left(  0\right)  }+H^{\prime},\nonumber\\
H^{\prime}\left(  \mathbf{q},\hat{k}_{z}\right)   &  =\alpha\left(
\mathbf{q}\right)  +\beta\left(  \mathbf{q}\right)  \hat{k}_{z}+\gamma\left(
\mathbf{q}\right)  \hat{k}_{z}^{2}.\nonumber
\end{align}
Then we have
\begin{align*}
\left\langle H_{\mathrm{eff}}\right\rangle _{mn}  &  =E_{m}\delta_{m,n}%
+\sum_{i,j=1}^{8}\left\langle F_{i}^{\left(  m\right)  }\left(  z\right)
\right\vert \left[  \alpha\left(  \mathbf{q}\right)  \right]  _{ij}%
|F_{j}^{\left(  n\right)  }\left(  z\right)  \rangle\\
&  +\langle F_{i}^{\left(  m\right)  }\left(  z\right)  |\left[  \beta\left(
\mathbf{q}\right)  \right]  _{ij}\hat{k}_{z}|F_{j}^{\left(  n\right)  }\left(
z\right)  \rangle\\
&  +\langle F_{i}^{\left(  m\right)  }\left(  z\right)  |\left[  \gamma\left(
\mathbf{q}\right)  \right]  _{ij}\hat{k}_{z}^{2}|F_{j}^{\left(  n\right)
}\left(  z\right)  \rangle.
\end{align*}

The contribution of the subbands other than the two lowest electron subbands
should also be considered in the reducing process, which can be done by using
L\"{o}wdin perturbation theory. We include the lowest 20 electron subbands and
54 highest hole subbands in the QW respectively and divide them into the
weakly coupled subsets $S_{1}$ and $S_{2}$. The set $S_{1}$ includes the two
lowest electron subbands $\left\vert \chi_{1}\right\rangle $ and $\left\vert
\chi_{2}\right\rangle $, the other subbands are included in the set $S_{2}$.
The Hamiltonian is reduced into set $S_{1}$ using the L\"{o}wdin perturbation
method,
\begin{equation}
H_{mm^{\prime}}^{\left(  2\right)  }=\frac{1}{2}\sum_{l}H_{ml}^{^{\prime}%
}H_{lm^{\prime}}^{^{\prime}}\left[  \frac{1}{E_{m}-E_{l}}+\frac{1}%
{E_{m^{\prime}}-E_{l}}\right]  ,\nonumber
\end{equation}
where the indices $m$ correspond to states in the set $A$, the indices $l$
correspond to states in the set $B$, and
\begin{equation}
H_{ml}^{^{\prime}}=\left\langle \psi_{m}\right\vert H^{\prime}\left\vert
\psi_{l}\right\rangle .\nonumber
\end{equation}

Finally we obtain the effective two-dimensional Hamiltonian $H_{\mathrm{eff}%
}^{4\times4}$ in the basis $\left\vert \chi_{1},\uparrow\right\rangle $,
$\left\vert \chi_{2},\downarrow\right\rangle $, $\left\vert \chi
_{1},\downarrow\right\rangle $, $\left\vert \chi_{2},\uparrow\right\rangle $:
\begin{equation}
\left(
\begin{array}
[c]{cccc}%
E_{1}+B_{1}\cdot k_{\parallel}^{2} & Ak_{+} & 0 & 0\\
Ak_{-} & E_{2}+B_{2}\cdot k_{\parallel}^{2} & 0 & 0\\
0 & 0 & E_{1}+B_{1}\cdot k_{\parallel}^{2} & Ak_{-}\\
0 & 0 & Ak_{+} & E_{2}+B_{2}\cdot k_{\parallel}^{2}%
\end{array}
\right)  ,\nonumber
\end{equation}
where%

\begin{subequations}
\begin{align}
E_{1}  &  =0.05351\mathrm{eV},\nonumber\\
B_{1}  &  =0.86571\mathrm{eV}\cdot\mathrm{nm}^{2},\nonumber\\
E_{2}  &  =0.14396\mathrm{eV},\nonumber\\
B_{2}  &  =0.67969\mathrm{eV}\cdot\mathrm{nm}^{2},\nonumber\\
A  &  =0.01041\mathrm{eV}\cdot\mathrm{nm}.\nonumber
\end{align}

In order to examine the validity of the 4-band Hamiltonian, we plot the band
structure of the GaAs/In$_{x}$Ga$_{1-x}$As/GaAs PQW calculated by the 4-band
model and compare it with the eight-band $\boldsymbol{k\cdot p}$ model (see
Fig. 2a in the manuscript). One can see clearly that the band structure
obtained from the 4-band model [the solid lines in Fig. 2a] is in good
agreement with that obtained from the eight-band model [the dashed lines in
Fig. 2a].

\section{APPENDIX F: VERIFICATION OF NON-TRIVIAL $Z_{2}$ TOPOLOGICAL
INVARIANT}

Topological insulators with dissipationless edge states and ordinary
insulators are distinguished by different $Z_{2}$ invariants. For 2D systems,
Fu and Kane \cite{Fu2} have shown that the $Z_{2}$ invariant can be determined
from the parity of the occupied band at the four time-reversal invariant
momenta in the Brillouin zone. The $Z_{2}$ invariant $\nu=0,1$, which
distinguishes the quantum spin-Hall phase in two dimensions, is given by%
\end{subequations}
\begin{align*}
\left(  -1\right)  ^{\nu}  &  =%
%TCIMACRO{\tprod \nolimits_{i}}%
%BeginExpansion
{\textstyle\prod\nolimits_{i}}
%EndExpansion
\delta_{i},\\
\delta_{i}  &  =%
%TCIMACRO{\tprod \nolimits_{m=1}^{N}}%
%BeginExpansion
{\textstyle\prod\nolimits_{m=1}^{N}}
%EndExpansion
\xi_{2m}\left(  \Gamma_{i}\right)  ,
\end{align*}
where $\xi_{2m}\left(  \Gamma_{i}\right)  =\pm1$ is the parity eigenvalue of
the $2m$th occupied energy band at the time-reversal invariant point
$\Gamma_{i}$, which shares the same eigenvalue $\xi_{2m}=\xi_{2m-1}$ with its
Kramer degenerate partner. The four time-reversal invariant points
$\Gamma_{i=\left(  n_{1}n_{2}\right)  }=\left(  n_{1}\mathbf{Q}_{1}%
+n_{2}\mathbf{Q}_{2}\right)  /2$, where $n_{1},n_{2}=0,1$. The calculated
parity eigenvalue of the $2m$th ($m=1,2,3,4,5,6$) occupied energy band at
$\Gamma_{i}$ are listed:%
\[%
\begin{tabular}
[c]{c|cccc}%
$m$ & $\xi_{2m}\left(  \Gamma_{00}\right)  $ & $\xi_{2m}\left(  \Gamma
_{01}\right)  $ & $\xi_{2m}\left(  \Gamma_{10}\right)  $ & $\xi_{2m}\left(
\Gamma_{11}\right)  $\\\hline
$1$ & $+$ & $-$ & $-$ & $-$\\
$2$ & $+$ & $+$ & $+$ & $+$\\
$3$ & $-$ & $-$ & $-$ & $-$\\
$4$ & $+$ & $+$ & $+$ & $+$\\
$5$ & $+$ & $+$ & $+$ & $+$\\
$6$ & $-$ & $-$ & $-$ & $-$%
\end{tabular}
.
\]

From the above calculation, we can confirm that the $Z_{2}$ invariant $\nu=1$
at the $2m$th ($m=2,5,6$) occupied band where the minigaps open, and the
system enters the TI phase and the dissipationless edge states appear.


\begin{thebibliography}{99}                                                                                               %


\bibitem {TQS1}D. J. Thouless, M. Kohmoto, M. P. Nightingale, and M. den Nijs,
\emph{Quantized Hall Conductance in a Two-Dimensional Periodic Potential},
Phys. Rev. Lett. \textbf{49}, 405 (1982).

\bibitem {TQS2}Q. Niu, D. J. Thouless, and Y. S. Wu, \emph{Quantized Hall
conductance as a topological invariant}, Phys. Rev. B \textbf{31} 3372 (1985).

\bibitem {TQS3}F. D. M. Haldane, \emph{Model for a Quantum Hall Effect without
Landau Levels: Condensed-Matter Realization of the "Parity Anomaly"}, Phys.
Rev. Lett. \textbf{61}, 2015 (1988).

\bibitem {RMP1}M. Z. Hasan and C. L. Kane, \emph{Colloquium: Topological
insulators}, Rev. Mod. Phys. \textbf{82}, 3045 (2010).

\bibitem {RMP2}X. L. Qi and S. C. Zhang, \emph{Topological insulators and
superconductors}, Rev. Mod. Phys. \textbf{83}, 1057 (2011).

\bibitem {Kane}C. L. Kane and E. J. Mele, \emph{Z$_{2}$ Topological Order and
the Quantum Spin Hall Effect}, Phys. Rev. Lett. \textbf{95}, 146802 (2005).

\bibitem {BHZ}B. A. Bernevig, T. L. Hughes, and S. C. Zhang, \emph{Quantum
Spin Hall Effect and Topological Phase Transition in HgTe Quantum Wells},
Science\textbf{314}, 1757 (2006).

\bibitem {Konig1}M. K\"{o}nig, S. Wiedmann, C. Br\"{u}ne, A. Roth, H. Buhmann,
L. W. Molenkamp, X. L. Qi, S. C. Zhang, \emph{Quantum Spin Hall Insulator
State in HgTe Quantum Wells}, Science \textbf{318}, 766 (2007).

\bibitem {Konig2}I.Knez, R. R. Du, and G. Sullivan, \emph{Evidence for Helical
Edge Modes in Inverted InAs/GaSb Quantum Wells}, Phys. Rev. Lett.
\textbf{107}, 136603 (2011).

\bibitem {Fu}L. Fu, C. L. Kane, and E. J. Mele, \emph{Topological Insulators
in Three Dimensions}, Phys. Rev. Lett. \textbf{98}, 106803 (2007);

\bibitem {Fu2}L. Fu and C. L. Kane, \emph{Topological insulators with
inversion symmetry}, Phys. Rev. B \textbf{76}, 045302 (2007).

\bibitem {Hsieh}D. Hsieh, D. Qian, L. Wray, Y. Xia, Y. S. Hor, R. J. Cava, and
M. Z. Hasan, \emph{A topological Dirac insulator in a quantum spin Hall
phase}, Nature \textbf{452}, 970 (2008).

\bibitem {ZXShen}Y. L. Chen, J. G. Analytis, J.-H. Chu, Z. K. Liu, S.-K. Mo,
X. L. Qi, H. J. Zhang, D. H. Lu, X. Dai, Z. Fang, S. C. Zhang, I. R. Fisher,
Z. Hussain, and Z.-X. Shen, \emph{Experimental Realization of a
Three-Dimensional Topological Insulator, Bi$_{2}$Te$_{3}$}, Science
\textbf{325}, 178 (2009).

\bibitem {Hasan}Y. Xia, D. Qian, D. Hsieh, L. Wray, A. Pal, H. Lin, A. Bansil,
D. Grauer, Y. S. Hor, R. J. Cava, and M. Z. Hasan, \emph{Observation of a
large-gap topological-insulator class with a single Dirac cone on the
surface}, Nature Phys. \textbf{5}, 398 (2009).

\bibitem {Lin}H. Lin, L. A. Wray, Y. Xia, S. Xu, S. Jia, R. J. Cava, A.
Bansil, and M. Z. Hasan, \emph{Half-Heusler ternary compounds as new
multifunctional experimental platforms for topological quantum phenomena},
Nature Mater. \textbf{9}, 546 (2010).

\bibitem {Franz}M. Franz, \emph{Topological insulators: Starting a new
family}, Nature Materials \textbf{9}, 536 (2010).

\bibitem {KYang}K. Yang, W. Setyawan, S. Wang, M. B. Nardelli, and S.
Curtarolo , \emph{A search model for topological insulators with
high-throughput robustness descriptors}, Nature Mater. \textbf{11}, 614 (2012).

\bibitem {Chadov}S. Chadov, X. L. Qi, J\"{u}rgen K\"{u}bler, G. H. Fecher, C.
Felser, and S. C. Zhang, \emph{Tunable multifunctional topological insulators
in ternary Heusler compounds}, Nature Mater. \textbf{9}, 541 (2010).

\bibitem {Heusler}D. Xiao, Y. Yao, W. Feng, J. Wen, W. Zhu, X. Q. Chen, G. M.
Stocks, and Z. Zhang, \emph{Half-Heusler Compounds as a New Class of
Three-Dimensional Topological Insulators}, Phys. Rev. Lett. \textbf{105},
096404 (2010).

\bibitem {GaAs}O. P. Sushkov and A. H. Castro Neto, \emph{Topological
Insulating States in Laterally Patterned Ordinary Semiconductors}, Phys. Rev.
Lett. \textbf{110}, 186601 (2013).

\bibitem {Oxide}D. Xiao, W. G. Zhu, Y. Ran, N. Nagaosa, and S. Okamoto,
\emph{Interface engineering of quantum Hall effects in digital transition
metal oxide heterostructures}, Nature Commun. \textbf{2}, 596 (2011).

\bibitem {Tinfilm}Y. Xu, B. H. Yan, H. J. Zhang, J. Wang, G. Xu, P. Z. Tang,
W. H. Duan, and S. C. Zhang, \emph{Large-Gap Quantum Spin Hall Insulators in
Tin Films}, Phys. Rev. Lett. \textbf{111}, 136804 (2013).

\bibitem {Chang1}J. Li and K. Chang, \emph{Electric field driven quantum phase
transition between band insulator and topological insulator}, Appl. Phys.
Lett. \textbf{95}, 222110 (2009).

\bibitem {Chang2}M. S. Miao, Q. Yan, C. G. Van de Walle, W. K. Lou, L. L. Li,
and K. Chang, \emph{Polarization-Driven Topological Insulator Transition in a
GaN/InN/GaN Quantum Well}, Phys. Rev. Lett. \textbf{109}, 186803 (2012).

\bibitem {Chang3}D. Zhang, W. K. Lou, M. S. Miao, S. C. Zhang, and K. Chang,
\emph{Interface-Induced Topological Insulator Transition in GaAs/Ge/GaAs
Quantum Wells}, Phys. Rev. Lett. \textbf{111}, 156402 (2013).

\bibitem {Franz2}J. Hu, J. Alicea, R. Q. Wu, and M. Franz, \emph{Giant
Topological Insulator Gap in Graphene with 5$d$ Adatoms}, Phys. Rev. Lett.
\textbf{109}, 266801 (2012).

\bibitem {Katsnelson}M. I. Katsnelson, F. Guinea, and M. A. H. Vozmediano,
\emph{In-plane magnetic textures at the surface of topological insulators},
EUROPHYS LETT 104, 17001 (2013).

\bibitem {Wilczek}F. Wilczek and A. Zee, \emph{Appearance of Gauge Structure
in Simple Dynamical Systems}, Phys. Rev. Lett. \textbf{52}, 2111 (1984).

\bibitem {CPSun}C. P. Sun and M. L. Ge, \emph{Generalizing Born-Oppenheimer
approximations and observable effects of an induced gauge field}, Phys. Rev. D
\textbf{41}, 1349 (1990).

\bibitem {Moore}J. E. Moore and L. Balents, \emph{Topological invariants of
time-reversal-invariant band structures}, Phys. Rev. B \textbf{75}, 121306(R) (2007).

\bibitem {Galitski}N. H. Lindner, G. Refael and V. Galitski, \emph{Floquet
topological insulator in semiconductor quantum wells}, Nature Phys.
\textbf{7}, 490 (2011).

\bibitem {Antidot}D. Weiss, M.L. Roukes, A. Menschig, P. Grambow, K. von
Klitzing, G. Weimann, \emph{Electron pinball and commensurate orbits in a
periodic array of scatterers}, Phys. Rev. Lett., \textbf{66}, 2790 (1991).

\bibitem {Antidot1}J. Eroms, M. Zitzlsperger, D. Weiss, J. H. Smet, C.
Albrecht, R. Fleischmann, M. Behet, J. DeBoeck, G. Borghs, \emph{Skipping
orbits and enhanced resistivity in large-diameter InAs/GaSb antidot lattices},
Phys. Rev. B \textbf{59}, 7829(R) (1999).

\bibitem {Antidot2-1}C. Albrecht, J. H. Smet, D. Weiss, K. von Klitzing, R.
Hennig, M. Langenbuch, M. Suhrke, U. R\H{o}ssler, V. Umansky, H. Schweizer,
\emph{Fermiology of Two-Dimensional Lateral Superlattices}, Phys. Rev. Lett.
\textbf{83}, 2234 (1999).

\bibitem {Antidot2-2}C. Albrecht, J. H. Smet, K. von Klitzing, D. Weiss, V.
Umansky, H. Schweizer, \emph{Evidence of Hofstadter's Fractal Energy Spectrum
in the Quantized Hall Conductance}, Phys. Rev. Lett. \textbf{86}, 147 (2001).

\bibitem {Antidot3-1}K. Bittkau, Ch. Menk, Ch. Heyn, D. Heitmann, and C. M.
Hu, \emph{Far-infrared photoconductivity of electrons in an array of
nanostructured antidots}, Phys. Rev. B \textbf{68}, 195303 (2003).

\bibitem {Antidot3-2}Z. Q. Yuan, C. L. Yang, R. R. Du, L. N. Pfeiffer, and K.
W. West, \emph{Microwave photoresistance of a high-mobility two-dimensional
electron gas in a triangular antidot lattice}, Phys. Rev. B \textbf{74},
075313 (2006).

\bibitem {Antidot4}C. H. Park and S. G. Louie, \emph{Making Massless Dirac
Fermions from a Patterned Two-dimensional Electron Gas}, Nano Lett.
\textbf{9}, 1793 (2009).

\bibitem {Antidot5}S. Wang, L. Z. Tan, W. Wang, S. G. Louie, and N. Lin,
\emph{Manipulation and Characterization of Aperiodical Graphene Structures
Created in a Two-Dimensional Electron Gas}, Phys. Rev. Lett. \textbf{113},
196803 (2014).

\bibitem {paraQW1-1}A. Saced\'{o}n, F. Gonz\'{a}lez-Sanz, E. Calleja, E.
Mu\~{n}oz, S. I. Molina, F. J. Pacheco, D. Ara\'{u}jo, R. Garc\'{\i}a, M.
Louren\c{c}o, Z. Yang, P. Kidd, and D. Dunstan, \emph{Design of InGaAs linear
graded buffer structures}, Appl. Phys. Lett. \textbf{66}, 3334 (1995).

\bibitem {paraQW1-2}J. Liang, Y. C. Chua, M. O. Manasreh, E. Marega, Jr., and
G. J. Salamo, \emph{Broad-band photoresponse from InAs quantum dots embedded
into InGaAs graded well}, IEEE Electron Device Letters \textbf{26}, 631 (2005).

\bibitem {paraQW2}N. Dai, G. A. Khodaparast, F. Brown, R. E. Doezema, S. J.
Chung, and M. B. Santos, \emph{Band offset determination in the strained-layer
InSb/Al$_{x}$In$_{1-x}$Sb system}, Appl. Phys. Lett. \textbf{76}, 3905 (2000).

\bibitem {MIM}K. Lai, W. Kundhikanjana, M. A. Kelly, Z. X. Shen, J. Shabani,
and M. Shayegan, \emph{Imaging of Coulomb-Driven Quantum Hall Edge States},
Phys. Rev. Lett. \textbf{107}, 176809 (2011).

\bibitem {ARPAS2013}J. A. Sobota, S.-L. Yang, A. F. Kemper, J. J. Lee, F. T.
Schmitt, W. Li, R. G. Moore, J. G. Analytis, I. R. Fisher, P. S. Kirchmann, T.
P. Devereaux, and Z. X. Shen, \emph{Direct Optical Coupling to an Unoccupied
Dirac Surface State in the Topological Insulator Bi$_{2}$Se$_{3}$}, Phys. Rev.
Lett. \textbf{111}, 136802 (2013).

\bibitem {ARPAS1999}Y. Ueda, A. Furuta, H. Okuda, M. Nakatake, H. Sato, H.
Namatame, M. Taniguchi, \emph{Photoemission and inverse-photoemission studies
of Bi$_{2}$Y$_{3}$(Y=S, Se, Te)semiconductors}, J. Electron Spectrosc. Relat.
Phenom. \textbf{101}, 677 (1999).

\bibitem {ARPAS2012}D. Niesner, Th. Fauster, S. V. Eremeev, T. V.
Menshchikova, Yu. M. Koroteev, A. P. Protogenov, E. V. Chulkov, O. E.
Tereshchenko, K. A. Kokh, O. Alekperov, A. Nadjafov, and N. Mamedov,
\emph{Unoccupied topological states on bismuth chalcogenides}, Phys. Rev. B
\textbf{86}, 205403 (2012).

\bibitem {Kono}W. D. Rice, J. Kono, S. Zybell, S. Winnerl, J. Bhattacharyya,
H. Schneider, M. Helm, B. Ewers, A. Chernikov, M. Koch, S. Chatterjee, G.
Khitrova, H. M. Gibbs, L. Schneebeli, B. Breddermann, M. Kira, and S. W. Koch,
\emph{Observation of Forbidden Exciton Transitions Mediated by Coulomb
Interactions in Photoexcited Semiconductor Quantum Wells}, Phys. Rev. Lett
\textbf{110}, 137404 (2013).

\bibitem {Giti}T. R. Merritt, M. A. Meeker, B. A. Magill, G. A. Khodaparast,
S. McGill, J. G. Tischler, S. G. Choi, and C. J. Palmstrom,
\emph{Photoluminescence lineshape and dynamics of localized excitonic
transitions in InAsP epitaxial layers}, J. Appl. Phys. \textbf{115}, 193503 (2014).

\bibitem {Heremans}Y. Zhang, and J. J. Heremans, \emph{Effects of
ferromagnetic nanopillars on spin coherence in an InGaAs quantum well}, Solid
State Commun. \textbf{177}, 36 (2014).

\bibitem {Lowdin}P. O. L\"{o}wdin, \emph{A Note on the Quantum-Mechanical
Perturbation Theory}, J. Chem. Phys. \textbf{19}, 1396 (1951).
\end{thebibliography}
\end{document}